\documentclass[preprint,12pt]{elsarticle}
\usepackage{ifpdf}
\usepackage{lineno}
\usepackage{graphicx,natbib,amssymb,lineno}
\usepackage{multirow}
\usepackage{multicol}

\begin{document}
\begin{frontmatter}
\title{Mechanism of luminescent enhancement in Ba$_{2}$GdNbO$_{6}$:Eu$^{3+}$ perovskite by Li$^{+}$ co-doping}

\author[ifse]{Marcos Vinicius dos Santos Rezende}
\author[phys_ufs] {Mario Ernesto Giroldo Val\'erio}
\author[phys_ufma]{Rodolpho Mouta}
\author[phys_ufma]{Eduardo Moraes Diniz}
\author[phys_ufma,berk,berk2]{Carlos William de Araujo Paschoal \corref{corresp}}
\ead{paschoal@ufma.br}
\cortext[corresp]{Corresponding author: Phone: +55 98 3272 8222; Fax: +55 98 3272 8204; alternative e-mails: paschoal.william@gmail.com; paschoal.william@berkeley.edu}

\address[ifse]{Departamento de F\'{\i}sica, Universidade Federal de Sergipe, 49500-000, Itabaiana-Se, Brazil}
\address[phys_ufs]{Departamento de F\'{\i}sica, Universidade Federal de Sergipe, 49100-000 S\~ao Crist\'ov\~ao-SE, Brasil.}
\address[phys_ufma]{Departamento de F\'{\i}sica, CCET, Universidade Federal do Maranh\~ao, 65085-580, S\~ao Lu\'{\i}s - MA, Brazil}
\address[berk]{Department of Materials Science and Engineering, University of California Berkeley, 94720-1760, Berkeley - CA, United States}
\address[berk2]{Department of Physics, University of California Berkeley, 94720-7300, Berkeley - CA, United States}

\begin{abstract}
We investigated the Li$^{+}$ ion incorporation in Ba$_{2}$GdNbO$_{6}$:Eu$^{3+}$ perovskite by atomistic simulations based on energy minimization. We predicted the most probable sites occupied by Eu$^{3+}$ and Li$^{+}$ ions and the related charge-compensation mechanisms involved into these substitutions. The results show that the Eu$^{3+}$ and Li$^{+}$ ions are incorporated mainly at the Gd$^{3+}$ site. In the Li$^{+}$ ion case, there is a charge compensation by  Nb$_{Gd}^{\bullet\bullet}$ antisite. The crystal field parameters and the transition levels for the Eu$^{3+}$ ion in the BGN:Eu$^{3+}$ were calculated with basis on the simulated local symmetry of the Eu$^{3+}$ site. The results show that the mechanism of luminescent properties enhancement is the symmetry distortion induced by the Li$^{+}$ co-doping.
\end{abstract}

\end{frontmatter}


\section{Introduction}



The $Li^+$ ion has been extensively introduced into different oxide hosts, such as: $YVO_4:Eu^{3+}$ \cite{YANG2008A}, $SrAl_{2}O_{4}:Eu^{3+}$ \cite{QIANG2009}, $Y_2O_3:Eu^{3+}$ \cite{BAE2005,BAE2003}, $Gd_2O_3:Eu^{3+}$ \cite{YI2004}, $SrZnO_{2}:Eu^{3+}$ \cite{LI2010} and perovskite compounds \cite{TIAN2003,YANG2011,LIU2005} to improve their luminescent properties. Usually, Li$^{+}$ ions act as co-activators in these compounds.  Moreover, several studies showed that the $Li^+$ ion addition affects positively the morphology of particles as well as the luminescent efficiency of oxide materials. The improved luminescence intensity can be originated from local crystal field symmetry breaking around the rare-earth ions by the $Li^+$ doping. Particularly, Eu$^{3+}$ incorporation into Ba$_2$GdNbO$_6$ (BGN) matrix modifies the luminescence spectrum due to the creation of emission centers, which generates specific red light \cite{YU2007}. Since Li$^{+}$ ion is very small, it can occupy any site in the BGN structure, either substitutionally , at the $Ba^{2+}$, $Gd^{3+}$ or $Nb^{5+}$ sites, or interstitially. In both cases, to compensate the charge, additional defects are created, which can modify the local crystal field symmetry around the RE ions.

Recently, Liu et al \cite{YU2007} reported a detailed investigation about the synthesis and luminescence characterization of the  BGN:Eu$^{3+}$,Dy$^{3+}$ and $Li^{+}$ co-doped BGN:Eu$^{3+}$,Dy$^{3+}$ samples. They observed that the co-doping enhances the emissions of  BGN: Eu$^{3+}$/Dy$^{3+}$ samples and related this enhancement to the charge compensation mechanism, which plays an important role in improving the luminescence efficiency of phosphors \cite{YI2004,JANG2007C,PANG2005B}.

In this paper we investigated the charge compensation mechanisms due to the $Li^{+}$ incorporation into BGN:Eu$^{3+}$ perovskite. We identified, by atomistic modelling and a simple overlap model, the most probable charge compensation mechanisms and the local symmetry breaking induced by Li co-doping, which is the mechanism that enhances the luminescent properties.

\section{Computational method}

To model the pristine BGN crystal, a standard lattice-energy minimization using the General Utility Lattice Program (GULP) code \cite{GALE1997,GALE2003,GALE2005} was performed. Buckingham pairwise potentials were assumed for all interionic interactions, which were described together with the electrostatic interaction in the form:

\begin{equation} \label{1}
U_{ij} (r_{ij} )=\frac{Z_{i} Z_{j} e^{2} }{r_{ij} } +A_{ij} \exp \left[\frac{-r_{ij} }{\rho _{ij} } \right]-\frac{B_{ij} }{r_{ij}^{6} }.
\end{equation}

In this equation, the first term describes the long-range electrostatic interaction between the ions of charge $Z_{i} e$ and $Z_{j} e$ separated by the distance $r_{ij} $; the second term models the Pauli short-range repulsion and the last term models the van der Walls attraction. To compute the Coulomb term, the Ewald summation \cite{TOSI1964}, which is standard in GULP code, was employed. The barium and oxygen ions were treated by the shell model \cite{DICK1958}. In this model is assumed that the \textit{i$^{th}$} ion is formed by a massless shell with charge $Y_{i} $ and a core with mass $m_{i} $ whose charge is $Z_{i} e-Y_{i} e$, with $Z_{i} $ being the valence of the \textit{i$^{th}$} ion. To obtain a finite ionic polarizability the core is connected to the shell by a harmonic spring, whose force constant is $k_{i} $.  Gd$^{3+}$, Eu$^{3+}$ and Li$^+$ ions were modeled using a rigid ion model due to theirs low polarizability.

The basic point defect energies were calculated using the Mott--Littleton method  that considers the point defect in the centre of a region, in which all interactions immediately surrounding it (region I) are treated explicitly, while a continuum approach is used for more distant regions from the defect (region IIb). These two regions are connected by another one region called IIa, in which ions are allowed to relax, but assuming that they are in harmonic potential wells. \cite{MOTT1938}. Typical region radii of 12 {\AA} (region I) and 16 {\AA} (region IIa) were adopted.

The electronic transitions of the Eu$^{3+}$ ion were calculated using the modified crystal field theory based on the Judd–-Ofelt theory \cite{judd,ofelt}. In this theory, the energies are related to the crystal field Hamiltonian ($H_{CF}$) by:
\begin{equation}\label{h_bqk}
  H_{CK}=\sum_{k,q} B_q^k C_q^k
\end{equation}
In this equation, the $C_q^k$ parameters describe the contribution of the dopant ion to the crystal field Hamiltonian, while the $B_q^k$ parameters describe the corresponding contribution of the Eu$^{3+}$ surrounding oxygen ions. To calculate the $B_q^k$ parameters the simple overlap (SO) model \cite{SOM} was employed. The initial data used in the SO model were the dopant ion position and the relaxed positions of the surrounding ions, which were obtained by the defect calculations. This method was employed with success to optical transitions for other materials \cite{marcos1,marcos2,marcos3}.

\section{Results}

\subsection{Potential adopted and basic defect calculations}
BGN crystallizes in a tetragonal distorted perovskite that derives from the cubic rock salt double perovskite structure. However, as the tetragonal distortion is small, BGN can be well described by the pseudocubic structure \cite{KHALAM2004,DIAS2006}. We assumed a previous complete set of potentials assumed to model the pristine BGN crystal \cite{bush,Pirovano2001,PASCHOAL2008}, which is listed in Table \ref{pot}. This potential set exhibits an excellent reliability to model the structural properties of BGN. Besides,  the dielectric properties of BGN were remarkably modeled by this potential data, which is a necessary condition for consistent defect calculation \cite{Catlow1975}. The good reliability of the potential set assumed can be checked comparing the error between calculated and experimental values for structural and dielectric properties of BGN, which are lower than 1\%. Table \ref{pot} also shows the used interactions for Eu$^{3+}$ \cite{PASCHOAL2008} and Li$^+$, which were took from \cite{bush,PASCHOAL2008}.

\begin{table}[htbp]
\caption{\label{pot}Short-range potentials parameters assumed to model pristine BGN crystal \cite{PASCHOAL2008,Pirovano2001,bush}.}
\begin{tabular}{llll}
\hline
\multicolumn{4}{c}{\bf Shell model interactions}\\ \hline
\multicolumn{4}{l}{Shell-Shell interactions}\\ \cline{1-4}
Interaction & $A/$eV & $\rho/$\AA & $C/$eV\AA$^6$ \\
\hline
Ba$^{2+} -$ O$^{2-}$ & 4818.4160    & 0.30670 & 0.00 \\
Nb$^{5+} -$ O$^{2-}$ & 1796.30    & 0.345980 & 0.00 \\
O$^{2-} -$ O$^{2-}$  & 25.41  & 0.69370 & 32.32 \\
\hline
\multicolumn{4}{l}{Core-Shell interactions} \\ \cline{1-3}
Ion & $k/$eV\AA$^{-2}$ & $Y/|e|$ &  \\
\hline
Ba$^{2+}$ &    34.05 &  1.83100 & \\
Nb$^{5+}$ &  1358.58 & -4.49700 & \\
O$^{2-}$  &    20.53 & -2.51300 & \\ \hline
\multicolumn{4}{c}{\bf Rigid ion model interactions}\\ \hline
Interaction & $A/$eV & $\rho/$\AA & $C/$eV\AA$^6$ \\ \hline
Gd$^{3+} - $ O$^{2-}$  &  1204.60   & 0.334137 & 0.00 \\
Eu$^{3+} - $ O$^{2-}$  &  1156.72   & 0.337617 & 0.00 \\
Li$^{+} -$ O$^{2-}$  & 426.48  & 0.3000 & 0,00 \\
\hline
\end{tabular}
\end{table}

\begin{figure}[!htbp]
  \centering
  \includegraphics[width=8cm,bb=0 300 600 800]{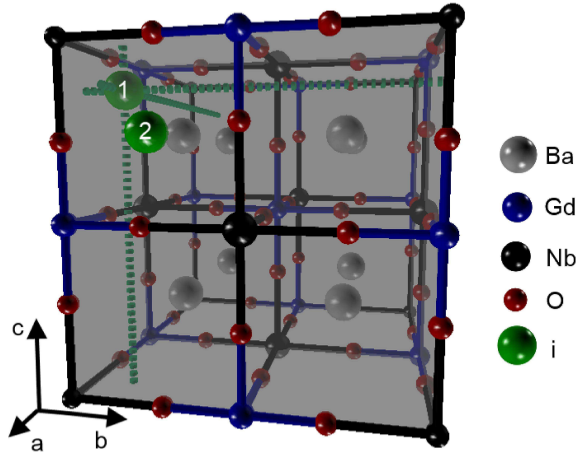}\\
  \caption{Pseudocubic unit cell of BGN showing the positions $i1=(\frac{1}{4},\frac{1}{4},0)$ and $i2=(\frac{1}{8},\frac{1}{8},\frac{1}{8})$ assumed as interstitial positions in this paper.}\label{figure1}
\end{figure}

We considered two possible interstitial positions to put each ion, intrinsic or extrinsic, in the pristine BGN crystal, namely: $i_1 = (\frac{1}{4}, \frac{1}{4}, 0)$ and $i_2 = (\frac{1}{8}, \frac{1}{8}, \frac{1}{8})$, as they are shown in Fig \ref{figure1}.  The position $i_{1}$ was in $xy$ plane in the square formed by the Nb and RE ions, while the position $i_{2}$ was in the diagonal between the Nb and RE ions. In the calculation, it was considered the formation energy of the most probable interstitial position. The formation energies of the basic point defects (vacancies and interstitials), as well as lattice energies, are given in Tables \ref{table1} and \ref{table2}, respectively. We used the Kr\"oger–Vink notation to label the defects.
\begin{table}[!htbp]
  \centering
  \caption{Formation energy of basic defects in the pristine BGN crystal. $i1=(\frac{1}{4},\frac{1}{4},0)$ and $i2=(\frac{1}{8},\frac{1}{8},\frac{1}{8})$ }\label{table1}
  \begin{tabular}{cc}
  \hline
    {\bf Defect} & {\bf Energy /eV} \\ \hline
  \multicolumn{2}{l}{{\bf Vacancies}}  \\ \hline

$V_{Ba}^{''}$                       & 20.90   \\
$V_{Gd}^{'''}$                      & 46.89   \\
$V_{Nb}^{'''''}$                    & 138.53  \\
$V_{O}^{\bullet \bullet}$           & 16.34   \\     \hline
\multicolumn{2}{l}{{\bf Interstitials at $i_1$ position}}  \\   \hline
 $Ba_{i1}^{\bullet \bullet}$                         & -9.90  \\
 $Gd_{i1}^{\bullet \bullet \bullet}$                 & -33.55  \\
 $Nb_{i1}^{\bullet \bullet \bullet \bullet \bullet}$ & -109.34 \\
 $O_{i1}^{''}$                                       & -7.40  \\
 $Li_{i1} ^{\bullet}$                                & -6.48 \\  \hline
 \multicolumn{2}{l}{{\bf Interstitials at $i_2$ position}}       \\   \hline
 $Ba_{i2}^{\bullet \bullet}$                         & -9.70   \\
 $Gd_{i2}^{\bullet \bullet \bullet}$                 & -31.87  \\
 $Nb_{i2}^{\bullet \bullet \bullet \bullet \bullet}$ & -106.28 \\
 $O_{i2}^{''}$                                       & -6.84   \\
 $Li_{i2} ^{\bullet}$                                & -5.31 \\  \hline
 \multicolumn{2}{l}{{\bf Substitution energies in BGN}}                \\   \hline
 $Nb_{Gd}^{\bullet \bullet}$         & -83.51 \\
 $Nb_{Ba}^{\bullet \bullet \bullet}$ & -92.45 \\
 $Ba_{Nb}^{'''}$                     & 116.45 \\
 $Ba_{Gd}^{'}$                       & 25.62  \\
 $Gd_{Ba}^{\bullet}$                 & -20.62 \\
 $Gd_{Nb}^{''}$                      & 88.49  \\
 $Eu_{Ba}^{\bullet}$                 & -20.52 \\
 $Eu_{Gd}$                           & 0.15 \\
 $Eu_{Nb}^{''}$                           & 88.65 \\
 $Li_{Ba}^{'}$                 & 12.90 \\
 $Li_{Gd}^{''}$                 & 36.89 \\
 $Li_{Nb}^{''''}$                 & 127.41 \\

  \hline
\end{tabular}

\end{table}

\begin{table}[!htbp]
  \centering
  \caption{Calculated lattice energies for the pristine BGN crystal and the start oxides.} \label{table2}
  \begin{tabular}{cc}
    \hline
    Compound & $E_{latt}$ / eV \\ \hline
    BGN & -297.63 \\
    $BaO$ & -32.18 \\
    $Gd_{2}O_{3}$ & -129.56 \\
    $Nb_{2}O_{5}$ & -323.72 \\
    $Li_{2}O$ & -30.51 \\
    $Eu_{2}O_{3}$ & -129.28 \\
    \hline
  \end{tabular}

\end{table}

The basic defects that involve Nb$^{5+}$ ions, i. e., Nb vacancies $(V_{Nb}^{'''''})$, Nb interstitials $(Nb_{i}^{\bullet \bullet \bullet \bullet \bullet})$ have higher absolute value of formation energies than those involving other ions ($Ba^{2+}$, $Gd^{3+}$ and $O^{2-}$). Besides, the more negative the antisite defect charge is, the more positive is its formation energy (and the opposite is also true). Such behaviours can be explained as follows: each ion has a contribution for the (negative) net lattice energy, which is mainly due to the electrostatic potential. The greater the modulus of its charge, the greater this contribution. Furthermore, because of the energy extensivity, by adding an ion to the crystal, the net lattice energy decreases; by removing it, the net lattice energy increases. Therefore, by creating a Nb vacancy, the contribution loss is greater than by creating any other vacancies, resulting in a more positive net lattice energy; when a Nb interstitial is created, the contribution gain is greater than that of any other interstitial. As for antisites, when we replace an ion by another one with lower charge, we reduce the contribution, increasing the net lattice energy. The opposite is also true: more positively charged antisites must have a more negative formation energy. That is also the reason why the antisite $Eu_{Gd}$, whose charge is null, has the lower energy of all. Thus, it is evident that our results are in complete agreement to what one should expect for the basic defect formation energies.

\subsection{Eu$^{3+}$ inclusion}
As the Eu$^{3+}$ ion is a trivalent lanthanide, it is expected it occupies substitutionally the Gd$^{3+}$ site without charge compensation. However, all other possible defect configurations  cannot be discarded, i. e., the incorporation into the $Ba^{2+}$ and Nb$^{5+}$ sites also need to be evaluated.  It is important to point out that the incorporation into the Ba$^{2+}$ and Nb$^{5+}$ sites generates more than one possible defects due to different charge compensation mechanisms. We listed all types of defects considered due to the Eu$^{3+}$ inclusion in the pristine BGN crystal and the charge compensation mechanisms associated to them in Table \ref{defects_Eu_Dy}. The results confirm that the most probable inclusion of Eu$^{3+}$ is into the Gd$^{3+}$ site.
\begin{table}[!htbp]
\tiny
\caption{\label{defects_Eu_Dy} Types of Defects considered due to the Eu$^{3+}$ inclusions in the pritine BGN crystal and solution energies by defect of each reaction in eV.}
\begin{tabular}{lclcc}
\hline
{\bf Site} & {\bf Charge compensation} & {\bf Reaction} & {\bf Energy} & \\ \hline
Gd$^{3+}$  & No charge compensation                                                     & $\frac{1}{2}Eu_2O_3+Gd^x_{Gd}\rightarrow\left(Eu_{Gd}^{x}\right)+\frac{1}{2}Gd_2O_3$                                                 & 0.01 \\ \hline
\multirow{7}{*}{Ba$^{2+}$} & Barium vacancies                                          & $Eu_2O_3+3Ba^x_{Ba}\rightarrow\left(2Eu_{Ba}^{\bullet}+V_{Ba}''\right)+3BaO$                                                            & 4.20  \\
                                & Oxygen interstitial                                       & $Eu_2O_3+2Ba^x_{Ba}\rightarrow\left(2Eu_{Ba}^{\bullet}+O_i''\right)+2BaO$                                                            & 5.49  \\
                                & Gadolinium vacancies                                      & $\frac{3}{2}Eu_2O_3+3Ba^x_{Ba}+Gd^x_{Gd}\rightarrow\left(3Eu_{Ba}^{\bullet}+V_{Gd}'''\right)+3BaO+\frac{1}{2}Gd_2O_3$                  & 4.48  \\
                                & Niobium vacancies                                         & $\frac{5}{2}Eu_2O_3+5Ba^x_{Ba}+Nb^x_{Nb}\rightarrow\left(5Eu_{Ba}^{\bullet}+V_{Nb}'''''\right)+5BaO+\frac{1}{2}Nb_2O_5$               & 6.06  \\
                                & Anti-site $\left(Ba_{Gd}'\right)$                         & $\frac{1}{2}Eu_2O_3+Ba^x_{Ba}+Gd^x_{Gd}\rightarrow\left(Eu_{Ba}^{\bullet}+Ba_{Gd}'\right)+\frac{1}{2}Gd_2O_3$                         & 2.48  \\
                                & Anti-site $\left(Gd_{Nb}''\right)$                        & $Eu_2O_3+2Ba^x_{Ba}+\frac{1}{2}Gd_2O_3\rightarrow\left(2Eu_{Ba}^{\bullet}+Gd_{Nb}''\right)+2BaO+\frac{1}{2}Nb_2O_5$                  & 5.10  \\
                                & Anti-site $\left(Ba_{Nb}'''\right)$                       & $\frac{3}{2}Eu_2O_3+3Ba^x_{Ba}+Nb^x_{Nb}\rightarrow\left(3Eu_{Ba}^{\bullet}+Ba_{Nb}'''\right)+2BaO+\frac{1}{2}Nb_2O_5$                 & 5.65  \\  \hline
\multirow{7}{*}{Nb$^{5+}$ } & Oxygen vacancies                                          & $\frac{1}{2}Eu_2O_3+O_O+Nb^x_{Nb}\rightarrow\left(Eu_{Nb}''+V_O^{\bullet\bullet}\right)+\frac{1}{2}Nb_2O_5$                          & 3.89  \\
                                & Barium interstitial                                       & $\frac{1}{2}Eu_2O_3+BaO+Nb^x_{Nb}\rightarrow\left(Eu_{Nb}''+Ba_i^{\bullet\bullet}\right)+\frac{1}{2}Nb_2O_5$                         & 6.95  \\
                                & Gadolinium interstitial                                   & $\frac{3}{2}Eu_2O_3+Gd_2O_3+3Nb^x_{Nb}\rightarrow\left(3Eu_{Nb}''+2Gd_i^{\bullet\bullet\bullet}\right)+\frac{3}{2}Nb_2O_5$           & 7.35  \\
                                & Niobium interstitial                                      & $\frac{5}{2}Eu_2O_3+5Nb^x_{Nb}\rightarrow\left(5Eu_{Nb}''+2Nb_i^{\bullet\bullet\bullet\bullet}\right)+\frac{3}{2}Nb_2O_5$            & 8.88  \\
                                & Anti-site $\left((Gd_{Ba}^{\bullet}\right)$               & $\frac{1}{2}Eu_2O_3+2Ba^x_{Ba}+Nb^x_{Nb}+Gd_2O_3\rightarrow\left(5Eu_{Nb}''+2Gd_{Ba}^{\bullet}\right)+2BaO+\frac{1}{2}Nb_2O_5$         & 5.13  \\
                                & Anti-site $\left((Nb_{Gd}^{\bullet\bullet}\right)$        & $\frac{1}{2}Eu_2O_3+Gd^x_{Gd}+Nb^x_{Nb}+\rightarrow\left(5Eu_{Nb}''+Nb_{Gd}^{\bullet\bullet}\right)+\frac{1}{2}Gd_2O_3$                & 2.50  \\
                                & Anti-site $\left((Nb_{Ba}^{\bullet\bullet\bullet}\right)$ & $\frac{3}{2}Eu_2O_3+Ba^x_{Ba}+3Nb^x_{Nb}+\rightarrow\left(3Eu_{Nb}''+2Nb_{Ba}^{\bullet\bullet\bullet}\right)+2BaO+\frac{1}{2}Nb_2O_5$  & 9.75  \\ \hline
\end{tabular}                                                                                                                                                                                                                  \end{table}                                                                                                                                                                                                                      
\subsection{Li$^{+}$ inclusion}

Considering the BGN crystalline structure, the $Li^{+}$ ion can be inserted into the $Ba^{2+}$, $Gd^{3+}$, $Nb^{5+}$ sites and in the most probable interstice ($i_1$). For all of them, there are more than one charge-compensating mechanism. We listed all type of defects considered due to the Li$^{+}$ inclusion into the pristine BGN crystal and the charge compensation mechanisms associated to them in Table \ref{defects_Li}. The solution energies (unbound) per defect are also shown.
\begin{table}[!htbp]
\tiny
\caption{\label{defects_Li} Types of Defects considered due to the Li$^{+}$ inclusion in the pristine BGN crystal and solution energies by defect of each reaction in eV.}
\begin{tabular}{lclc}
\hline
{\bf Site} & {\bf Charge compensation} & Reaction & Energy \\ \hline
\multirow{7}{*}{Gd$^{3+}$}  & Oxygen vacancies                                          & $\frac{1}{2}Li_2O+O^x_O+Gd^x_{Gd}\rightarrow\left(Li_{Gd}''+V_O^{\bullet\bullet}\right)+\frac{1}{2}Gd_2O_3$                                    & 1.85  \\
                                & Barium interstitial                                       & $\frac{1}{2}Li_2O+BaO+Gd^x_{Gd}\rightarrow\left(Li_{Gd}''+Ba_i^{\bullet\bullet}\right)+\frac{1}{2}Gd_2O_3$                                   & 4.82  \\
                                & Gadolinium interstitial                                   & $\frac{3}{2}Li_2O+3Gd^x_{Gd}\rightarrow\left(3Li_{Gd}''+2Gd_i^{\bullet\bullet}\right)+\frac{3}{2}Gd_2O_3$                                    & 4.91  \\
                                & Niobium interstitial                                      & $\frac{5}{2}Li_2O+5Gd^x_{Gd}+Nb_2O_5\rightarrow\left(5Li_{Gd}''+2Nb_i^{\bullet\bullet\bullet\bullet\bullet}\right)+\frac{5}{2}Gd_2O_3$                          & 5.98  \\
                                & Anti-site $\left(Gd_{Ba}^{\bullet}\right)$                & $\frac{1}{2}Li_2O+\frac{1}{2}Gd_2O_3+2Ba^x_{Ba}+Gd^x_{Gd}\rightarrow\left(Li_{Gd}''+2Gd_{Ba}^{\bullet}\right)+2BaO$                            & 10.65  \\
                                & Anti-site $\left(Nb_{Gd}^{\bullet\bullet}\right)$         & $\frac{1}{2}Li_2O+2Gd^x_{Gd}+\frac{1}{2}Nb_2O_5\rightarrow\left(Li_{Gd}''+Nb_{Gd}^{\bullet\bullet}\right)+\frac{1}{2}Gd_2O_3$                           & 0.47  \\
                                & Anti-site $\left(Nb_{Ba}^{\bullet\bullet\bullet}\right)$  & $\frac{3}{2}Li_2O+3Gd^x_{Gd}+2Ba^x_{Ba}+Nb_2O_5\rightarrow\left(3Li_{Gd}''+2Nb_{Ba}^{\bullet\bullet\bullet}\right)+2BaO+\frac{3}{2}Gd_2O_3$    & 7.31  \\ \hline
\multirow{7}{*}{Ba$^{2+}$ } & Oxygen vacancies                                          & $Li_2O+O^x_O+Ba^x_{Ba}                            \rightarrow  \left(2Li_{Ba}'+V_O^{\bullet\bullet}\right)+2BaO$                               & 2.76  \\
                                & Barium interstitial                                       & $Li_2O+2Ba^x_{Ba}                               \rightarrow  \left(Li_{Ba}'+Ba_i^{\bullet\bullet}\right)+BaO$                                & 4.74  \\
                                & Gadolinium interstitial                                   & $\frac{1}{2}Li_2O+3Ba^x_{Ba}+\frac{1}{2}Gd_2O_3   \rightarrow  \left(3Li_{Ba}'+Gd_i^{\bullet\bullet\bullet}\right)+3BaO$                       & 4.79  \\
                                & Niobium interstitial                                      & $\frac{5}{2}Li_2O+5Ba^x_{Ba}+\frac{1}{2}Nb_2O_5     \rightarrow  \left(5Li_{Ba}'+Nb_i^{\bullet\bullet\bullet\bullet\bullet}\right)+5BaO$         & 5.40  \\
                                & Anti-site $\left(Gd_{Ba}^{\bullet}\right)$                & $\frac{1}{2}Li_2O+2Ba^x_{Ba}+\frac{1}{2}Gd_2O_3   \rightarrow  \left(Li_{Ba}'+Gd_{Ba}^{\bullet}\right)+2BaO$                                   & 3.98 \\
                                & Anti-site $\left(Nb_{Gd}^{\bullet\bullet}\right)$         & $Li_2O+2Ba^x_{Ba}+Gd^x_{Gd}+\frac{1}{2}Nb_2O_5      \rightarrow  \left(2Li_{Ba}'+Nb_{Gd}^{\bullet\bullet}\right)+2BaO+\frac{1}{2}Gd_2O_3$        & 1.84  \\
                                & Anti-site $\left(Nb_{Ba}^{\bullet\bullet\bullet}\right)$  & $\frac{3}{2}Li_2O+4Ba^x_{Ba}+\frac{1}{2}Nb_2O_5   \rightarrow  \left(3Li_{Ba}'+Nb_{Ba}^{\bullet\bullet\bullet}\right)+4BaO$                    & 2.48  \\   \hline

\multirow{7}{*}{Nb$^{5+}$ } & Oxygen vacancies                                          & $\frac{1}{2}Li_2O_2+O^x_O+Nb^x_{Nb}                             \rightarrow  \left( Li_{Nb}''''  +  2V_O^{\bullet\bullet}\right)+\frac{1}{2}Nb_2O_5$                            & 4.50    \\
                                & Barium interstitial                                       & $\frac{1}{2}Li_2O_2+BaO+Nb_{Nb}                            \rightarrow  \left( Li_{Nb}''''  +  2Ba_i^{\bullet\bullet}\right)+\frac{1}{2}Nb_2O_5 $                          & 8.46    \\
                                & Gadolinium interstitial                                   & $\frac{3}{2}Li_2O+3Nb^x_{Nb}+2Gd_2O_3                         \rightarrow  \left(3Li_{Nb}''''  +  4Gd_i^{\bullet\bullet\bullet}\right)+\frac{3}{2}Nb_2O_5 $                   & 9.62    \\
                                & Niobium interstitial                                      & $\frac{5}{2}Li_2O+5Nb_{Nb}                                  \rightarrow  \left(5Li_{Nb}''''  +  4Nb_i^{\bullet\bullet\bullet\bullet\bullet}\right)+\frac{1}{2}Nb_2O_5 $     & 12.68   \\
                                & Anti-site $\left(Gd_{Ba}^{\bullet}\right)$                & $\frac{1}{2}Li_2O+ 4Ba^x_{Ba}+\frac{1}{2}Gd_2O_3                \rightarrow  \left( Li_{Nb}''''  +  4Gd_{Ba}^{\bullet}\right)+4BaO+\frac{1}{2}Nb_2O_5 $                         & 5.74    \\
                                & Anti-site $\left(Nb_{Gd}^{\bullet\bullet}\right)$         & $\frac{1}{2}Li_2O+ Nb^x_{Nb}+2Gd^x_{Gd}+\frac{1}{2}Nb_2O_5      \rightarrow  \left( Li_{Nb}''''  +  2Nb_{Gd}^{\bullet\bullet}\right)+Gd_2O_3  $                                & 2.65    \\
                                & Anti-site $\left(Nb_{Ba}^{\bullet\bullet\bullet}\right)$  & $\frac{3}{2}Li_2O+3Nb^x_{Nb}+4Ba^x_{Ba}+\frac{1}{2}Nb_2O_5      \rightarrow  \left(3Li_{Nb}''''  +  4Nb_{Ba}^{\bullet\bullet\bullet}\right)+4BaO $                              & 13.05   \\ \hline
\multirow{7}{*}{Interstitial}   & Barium vacancies                         & $Li_2O+Ba^x_{Ba}                          \rightarrow  \left( 2Li_{i}^{\bullet} + V_{Ba}'' \right)+BaO  $                          & 2.09   \\
                                & Oxygem interstitial                      & $Li_2O                                    \rightarrow  \left( 2Li_{i}^{\bullet} + O_{i}''\right) $                                 & 3.38   \\
                                & Gadolinium vacancies                     & $\frac{3}{2}Li_2O+Gd^x_{Gd}               \rightarrow  \left(3Li_{i}^{\bullet}  + V_{Gd}'''\right)+\frac{1}{2}Gd_2O_3 $           & 2.11   \\
                                & Niobium vacancies                        & $\frac{5}{2}Li_2O+Nb^x_{Nb}               \rightarrow  \left(5Li_{i}^{\bullet}  +  V_{Nb}''''' \right)+\frac{1}{2}Nb_2O_5 $        & 3.42   \\
                                & Anti-site $\left(Ba_{Gd}^{'}\right)$     & $\frac{1}{2}Li_2O+ BaO+Gd^x_{Gd}          \rightarrow  \left( Li_{i}^{\bullet}  +  Ba_{Gd}' \right)+\frac{1}{2}Gd_2O_3  $          & 0.90   \\
                                & Anti-site $\left(Gd_{Nb}^{''}\right)$    & $Li_2O+\frac{1}{2}Gd_2O_3+Nb^x_{Nb}      \rightarrow  \left(2Li_{i}^{\bullet}  +  Gd_{Nb}''\right)+\frac{1}{2}Nb_2O_5  $         & 4.48   \\
                                & Anti-site $\left(Ba_{Nb}^{'''}\right)$   & $\frac{3}{2}Li_2O+BaO+Nb^x_{Nb}          \rightarrow  \left(3Li_{i}^{\bullet}  +  Ba_{Nb}'''\right)+\frac{1}{2}Nb_2O_5 $          & 3.27   \\ \hline
\end{tabular}                                                                                                                                                                                                                                                                                
\end{table}

According to the results, there is an energetic preference for the Li$^+$ incorporation into the Gd$^{3+}$ site compensated by $Nb_{Gd}^{\bullet\bullet}$ antisite, whose solution energy  is 0.47 eV. The following energetic preferences are: the incorporation into the interstitial site $i_1$, compensated by $Ba'_{Gd}$ antisite , with a solution energy of 0.90 eV; the Ba$^{2+}$ site, compensated by $Nb_{Gd}^{\bullet\bullet}$ antisite, with a solution energy of 1.84 eV; and into the Nb$^{5+}$  site, compensated by oxygen vacancies, with an energy of 1.85 eV. 

It is important to point out that we calculated the bound solution energies. The unbound solution assumes that there is no interaction between dopant and charge-compensating defect, while in the bound solution, the calculations are carried out for a configuration consisting of the dopant and charge-compensating defects in vicinity positions, meaning that the energies include the contribution of the binding energy of the defect.  The reason for including the bound solution energies was to perform a careful analysis of the relaxed configurations, which revealed the large distortions in the surrounding lattice caused by the Li$^{+}$ co-dopant and the charge-compensating defect required to keep the total charge neutral close to the Eu$^{3+}$ ion. The presence of a charge-compensating defect near the active centre can change the local symmetry of the active centre and, therefore, change the luminescence efficiency of phosphors.

\subsection{Simultaneous Eu$^{3+}$ and Li$^+$ inclusions}
To explain how the Li$^+$ co-doping enhances the luminescence efficiency of BGN:Eu$^{3+}$, as reported by Yu et al. \cite{YU2007}, we calculated the incorporation of Eu$^{3+}$ and Li$^+$ simultaneously into BGN crystal. According to Yu et al,  Li$^+$ co-doping helps to incorporate Eu$^{3+}$ and Dy$^{3+}$ into lattice sites by increasing the crystallinity. We considered only those configurations more favorable to incorporate the dopants and co-dopants in the BGN matrix, with basis on the previous results showed in Tables \ref{defects_Eu_Dy} and \ref{defects_Li}. Besides, by considering different charge compensation mechanisms, we also tested possible no symmetry configuration  of dopant in the matrix. Thus, we considered the Eu$^{3+}$ ion as dopant and the Li$^+$ ion as co-dopant, and we assumed that there exists a binding energy among both and the charge compensation defect. The reaction that describes these substitutions is:
\begin{equation}
  \frac{1}{2}Eu_2O_3+3Gd_{Gd}+\frac{1}{2}Li_2O+\frac{1}{2}Nb_2O_5 \rightarrow \left( Eu_{Gd}+Li_{Gd}''+Nb_{Gd}^{\bullet\bullet} \right) + \frac{1}{2}Gd_2O_3 \\
\end{equation}

For the bounded defect, the dopant, co-dopant and charge mechanism can be arranged in six non-equivalent arrangements (C1 to C6), as it is shown in Table \ref{configs}, which also lists the calculated defect formation and solution energies of the simultaneous Eu$^{3+}$ and Li$^+$ incorporations into the pristine BGN crystal for each one of the six no equivalent arrangements. According to the results there is a small difference between all configurations. The C6 configuration exhibits the lower energy value, being the most probable configuration.
\begin{table}[!htbp]
\tiny
  \centering
\begin{tabular}{cccccc}
  \hline
  {\bf Configuration} & \multicolumn{3}{c}{{\bf Ion positions}} & {\bf Formation energy / eV}	&  {\bf Solution energy / eV} \\ \hline
C1 &	Eu $\rightarrow (0,0,0)                     $ & Li $\rightarrow (\frac{1}{2},0,\frac{1}{2}) $ & Nb $\rightarrow  (1,0,0)                     $ &  -47.42  &  0.00   \\
C2 &	Eu $\rightarrow (1,0,0)                     $ & Li $\rightarrow (0,0,0)                     $ & Nb $\rightarrow  (\frac{1}{2},0,\frac{1}{2}) $ &  -47.35	&  0.02   \\
C3 &	Eu $\rightarrow (\frac{1}{2},0,\frac{1}{2}) $ & Li $\rightarrow (1,0,0)                     $ & Nb $\rightarrow  (0,0,0)                      $ &  -47.44	&  -0.01  \\
C4 &	Eu $\rightarrow (0,0,0)                     $ & Li $\rightarrow (\frac{1}{2},0,\frac{1}{2}) $ & Nb $\rightarrow  (1,0,1)                      $ &  -47.17	&  0.08   \\
C5 &	Eu $\rightarrow (1,0,1)                     $ & Li $\rightarrow (0,0,0)                     $ & Nb $\rightarrow  (\frac{1}{2},0,\frac{1}{2}) $ &  -47.27	&  0.05   \\
C6 &	Eu $\rightarrow (\frac{1}{2},0,\frac{1}{2}) $ & Li $\rightarrow (1,0,1)                     $ & Nb $\rightarrow  (0,0,0)                     $ &  -47.97	&  -0.19  \\ \hline
  \hline
\end{tabular}
  \caption{Possible arrangements of the dopant, co-dopant and charge compensation defect in pristine BGN crystal and respective formation and solution energies. }\label{configs}
\end{table}

As previously mentioned, the Li$^+$ ion co-doping induces lattice distortions in the BGN:Eu$^{3+}$ crystal, and this distortion results in a different crystalline field around the active ion. From the defect calculations we obtained the relaxed positions of Li$^+$ and Eu$^{3+}$ ions and the respective surrounding lattice ions. We used these positions as initial data to calculate the crystal field parameters $B_q^k$, which are shown in Table \ref{sete} for BGN:Eu$^{3+}$ pure and Li co-doped. The $B_q^k$ parameters provide an unambiguous indication of the local symmetry of the optically active centre in the material and helped to explain the observed system optical activity. The change in local symmetry around the Eu$^{3+}$ ion is in agreement with the variation in the intensity of the electric dipole and magnetic dipole transitions \cite{Reisfeld,Blasse}. For the non co-doped BGN:Eu$^{3+}$, vanishing $B_0^k$ values indicate that the site occupied by the Eu$^{3+}$ ion has high symmetry. Thus, it is expected that the magnetic dipole transition $^5D_0 \rightarrow ^7F_1$ of Eu$^{3+}$  dominates the BGN:Eu$^{3+}$ emission spectra, which is in excellent agreement with the emission spectra reported by Yu et al. \cite{YU2007}.

In the  BGN:Eu$^{3+}$,Li$^+$ crystal, all $B_q^k$ parameters have a non-zero value, which indicates the Eu$^{3+}$ site symmetry is rather low. The low symmetry is caused by a large deformation resulting from the incorporation of Eu$^{3+}$ into the lattice, the presence of the Li$^+$ co-dopant and the $Nb_{Gd}^{\bullet\bullet}$ antisite, all close to the Eu$^{3+}$ ion. This low symmetry implies that the co-doped crystal has more effective luminescence emission than the pure BGN:Eu$^{3+}$ crystal. 

The luminescence intensity ratio ($R$) of $^5D_0\rightarrow ^7F_1$ and $^5D_0\rightarrow ^7F_2$ transitions, also called  asymmetry ratio, is widely used as an indication of the degree of asymmetry in the vicinity of Eu$^{3+}$ ions \cite{Reisfeld2004}. From a comparison of the emission spectra of BGN:Eu$^{3+}$ and BGN:Eu$^{3+}$,Li$^+$ crystals, reported by Yu et al. \cite{YU2007}, it was noted that the ratio $R$ changed due to the Li$^+$ co-doping. The greater intensity ratio corresponds to a more distorted or asymmetric Eu$^{3+}$ site. In BGN:Eu$^{3+}$ crystal, the asymmetry ratio $R$($\sim$2.44) is lower than those for BGN:Eu$^{3+}$,Li$^+$ R ($\sim$4.06), suggesting that the symmetry of the Eu$^{3+}$ in co-doped BGN:Eu$^{3+}$,Li$^+$ crystals is higher than in BGN:Eu$^{3+}$. The $B_q^k$ parameters, showed in Table \ref{sete} confirm a decreasing of local symmetry of Eu$^{3+}$ ion caused by Li$^{+}$ incorporation.

\begin{table}[!htbp]
  \centering
   \caption{$B_q^k$ values for Eu$^{3+}$ calculated with basis on the relaxed lattice of doped BGN:Eu$^{3+}$.}\label{sete}
  \begin{tabular}{ccc}
     \hline
\multirow{2}{*}{$B^k_q$} & BGN:Eu$^{3+}$	& BGN:Eu$^{3+}$,Li$^{+}$      \\   \cline{2-3}
$B^2_0$	& 0	       &  -51.17     \\
$B^2_1$	& 0	       &  -1.03      \\
$B^2_2$	& 0	       &  -62.67     \\   \hline
$B^4_0$	& 3058.14	 & 2922.39     \\
$B^4_1$	& 0	       &  -0.91      \\
$B^4_2$	& 0	       &  75.64      \\
$B^4_3$	& 0	       &  -2.42      \\
$B^4_4$	& 1292.30	 & 1275.37     \\  \hline
$B^6_0$	& 609.85	 & 538.00      \\
$B^6_1$	& 0	       &  -11.64     \\
$B^6_2$	& 0	       &  -29.41     \\
$B^6_3$	& 0	       &  5.08       \\
$B^6_4$	& -1140.93 & 	-1092.43   \\
$B^6_5$	& 0	       &  -2.66      \\
$B^6_6$	& 0	       &  -43.62     \\

     \hline
   \end{tabular}

\end{table}

	Table \ref{oito} shows the energy transition predicted for the Eu$^{3+}$ ion in BGN:Eu$^{3+}$ and BGN:Eu$^{3+}$,Li$^+$ crystals, using the crystal field parameters ($B_q^k$ values) from Table \ref{sete}. The experimental data  are also shown for comparison. It is important to point out that the calculated values are predictions using only the $B_q^k$ values obtained with the relaxed lattice surrounding the dopant, without any necessary previous knowledge of the spectra of the real system. The percentage difference between the predicted and the experimental results for the $^5D_0\rightarrow ^7F_1$ and $^5D_0\rightarrow ^7F_2$ transition energies  are lower than 3\% for both crystals, which shows the reliability of this method to perform this kind of calculation. Another important aspect is the comparison between total splitting of the $^5D_0\rightarrow ^7F_j (j=1,2,3)$ transitions in BGN:Eu$^{3+}$ and BGN:Eu$^{3+}$,Li$^+$. The total splitting in $^5D_0\rightarrow ^7F_1$ in the BGN:Eu$^{3+}$ is predicted to be 0.1 nm, while in the BGN:Eu$^{3+}$,Li$^+$  is 1.7 nm. On the other hand, the predicted total splitting of the $^5D_0\rightarrow ^7F_2$ transitions is  11.4 nm in the BGN:Eu$^{3+}$ crystal, almost the same to that calculated in Li$^+$ co-doped, whose splitting is 12.0 nm. The predicted total splitting of the $^5D_0\rightarrow ^7F_3$ transitions is  10.4 nm and 9.3 nm in in the BGN:Eu$^{3+}$ crystal and Li$^+$ co-doped, respectively. In general, it is usually observed an increasing of total splitting in the Li$^+$ co-doped crystal. The Li$^+$ ion incorporation into BGN:Eu$^{3+}$ crystals also subtly influences the energy level shift and splitting of Eu$^{3+}$ ion, as can be observed in Table \ref{oito}.  Similar results were obtained for ZnS:Tm$^{3+}$,Li$^{+}$ \cite{Stambouli1996}.
\begin{table}[!htbp]
  \centering
   \caption{Transition Energies of Eu$^{3+}$ in the BGN:Eu$^{3+}$ and BGN:Eu$^{3+}$,Li$^{+}$ crystals.}\label{oito}
  \begin{tabular}{cccc}
    \hline
            & \multicolumn{2}{c}{BGN:Eu$^{3+}$}	& BGN:Eu$^{3+}$,Li$^+$         \\     \cline{2-4}
Transition	& Energy Exp. (nm)	& Energy Cal. (nm)	& Energy Cal. (nm) \\ \hline
\multirow{1}{*}{$^5D_0\rightarrow ^7F_0$}	&   -	                  & 565.8    & 565.9   \\   \hline
\multirow{3}{*}{$^5D_0\rightarrow ^7F_1$}	&                         &	577.5    & 577.2   \\
                                            &      593                & 577.6	 & 577.3   \\
                                            &                         &          & 578.5   \\   \hline
\multirow{5}{*}{$^5D_0\rightarrow ^7F_2$}	&                         & 592.7    &  592.9   \\
                                            &  	613                   & 594.1    &  594.7   \\
                                            &   640                   & 602.2    &  603.0   \\
                                            &                         & 604.1    &  603.2   \\
                                            &                         &	         &  604.9   \\   \hline
\multirow{7}{*}{$^5D_0\rightarrow ^7F_3$}	&                         &          & 631.4    \\
                                            &                         & 631.2    & 635.2    \\
                                            &                         & 635.7    & 636.6    \\
                                            &            656          & 636.4    & 637.1    \\
                                            &                         & 640.2    & 639.2    \\
                                            &                         & 641.6    & 640.7    \\
                                            &                         &          & 641.3    \\
    \hline
  \end{tabular}

\end{table}

Additionally, the creation of Nb$_{Gd}^{\bullet \bullet}$ defects needed to compensate the charge due to the Li$^+$ incorporation also could effectively promote the crystallinity leading to higher oscillating strengths for the optical transitions \cite{YI2004,Misbra1992}. Nb$_{Gd}^{\bullet \bullet}$ defects also might act as a sensitizer for the effective energy transfer due to the strong mixing of charge transfer states and, therefore, the enhanced Eu$^{3+}$ emission intensity. The charge compensation defects, generated by the Li$^+$  incorporation, positively could be related to the morphology and grain size change of particles in BGN:Eu$^{3+}$,Li$^+$ crystals as reported by Yu et al. work \cite{YU2007}.

\section{Conclusions}
We successfully modeled the defects induced by Eu$^{3+}$ and Li$^{+}$ dopants in Ba$_2$GdNbO$_6$ perovskite. The calculations indicate that the Eu$^{3+}$ and Li$^{+}$ ions preferentially are included into the Gd site, being the Li$^{+}$ incorporation compensated by the Nb$_{Gd}^{\bullet\bullet}$ antisite. The crystal field parameters $B^k_q$ and the transition energies were calculated for Eu$^{3+}$ in the BGN:Eu$^{3+}$ and BGN:Eu$^{3+}$,Li$^{+}$ crystals. In the BGN:Eu, a vanishing $B^2_0$ value indicates that the site occupied by the Eu$^{3+}$ ion is an inversion centre. In the other hand, in the BGN:Eu$^{3+}$,Li$^+$ are observed the non-zero values of $B^k_q$, which suggest a low symmetry site. An excellent agreement between the predicted and experimental values of the transition energies was obtained showing the reliability of method used, as well as explaining the enhancement mechanism of the luminescence properties in BGN:Eu$^{3+}$ by Li$^{+}$ co-doping.

\section{Acknowledgement}
This work was partially supported by the Brazilian funding agencies CAPES, CNPq, FAPITEC and FAPEMA.  C. W. A. Paschoal acknowledges R. Ramesh for all support at Univ. California Berkeley. The authors thank Dr. Julian Gale for allowing us to use the GULP code.

%
%
%
%


\end{document}